\documentstyle[preprint,aps]{revtex}
\begin{document}
\draft
\title{Conductivity of Metallic Si:B Near the 
Metal-Insulator Transition: Comparison between
Unstressed and Uniaxially Stressed Samples}
\author{S. Bogdanovich and M. P. Sarachik}
\address{Physics Department, City College of the City University of 
New York, New York, New York 10031}
\author{R. N. Bhatt}
\address{Department of Electrical Engineering, Princeton University, 
Princeton, New Jersey 08544-5263}
\date{\today}
\maketitle
\begin{abstract}
The low-temperature dc conductivities of barely metallic samples of p-type 
Si:B are compared for a series of samples with different dopant concentrations, 
n, in the absence of stress (cubic symmetry), and for a single sample driven 
from the metallic into the insulating phase by uniaxial compression, S.  For 
all values of temperature and stress, the conductivity of the stressed sample 
collapses onto a single universal scaling curve, 
$\sigma (S, T) = \sigma_0 (\Delta S/S_c)^\mu G[T/T^*(S)]$, with 
$T^* \propto (\Delta S)^{z\nu}$.  The scaling fit indicates that the 
conductivity of Si:B is $\propto T^{1/2}$ in the critical 
range.  Our data yield a critical conductivity exponent 
$\mu = 1.6$, considerably larger than the value reported in earlier 
experiments where the transition was crossed by varying the dopant 
concentration.  The larger exponent is based on data in a narrow range 
of stress near the critical value within which scaling holds.  
We show explicitly that the temperature dependences of the 
conductivity of stressed and unstressed Si:B are different,     
suggesting that a direct comparison of the critical behavior and 
critical exponents for stress-tuned and concentration-tuned transitions 
may not be warranted.
\end{abstract}

\section{Introduction}
\label{intro}

A continuous metal-insulator transition in the limit of zero temperature 
has been demonstrated over the past two decades since the pioneering results of
Rosenbaum{\it et al.}\cite{Rosen1} in a wide variety of disordered
electronic systems, including uncompensated and compensated doped 
semiconductors, amorphous metal-insulator mixtures, and magnetic 
semiconductors. The region near the transition 
has been studied by tuning through the transition using the 
standard method of varying the concentration of one of the constituents
\cite{Rosen1,Bishop,Thomas,castner,holcomb,ourSi:B,sarachikSi:P,stupp,shlimak,itoh}
by uniaxial stress\cite{paalanen,ourPRL},
using a magnetic field\cite{vonM} to vary the critical point, or using 
persistent photoconductivity to vary the doping in shallow levels.
\cite{katsumoto}.

In metal-semiconductor mixtures and compensated semiconductors (including the 
persistent photoconductor, doped $Al_xGa_{1-x}As$\cite{katsumoto}) the onset of 
the conductivity is found to be well described by a particularly simple form
in the metallic phase:

\begin{equation}
\sigma(t,T) = \sigma(t,0) + B T^{1/2}   
\label{equation}
\end{equation}
where $\sigma(t,0) = A(t-t_c)^\mu$ is the 
zero temperature conductivity, the critical 
conductivity exponent $\mu \approx 1$, and the coefficient $B$ of the 
temperature-dependent term is independent of the tuning parameter $t$ 
(the metal fraction, dopant concentration, stress,
magnetic field, photo-induced carrier density etc.)
as it approaches the critical value $t_c$ at
the metal-insulator transition.
Measurements of the conductivity for 
different values of $t$ 
are thus found to yield a set of parallel straight lines when plotted against $T^{1/2}$. 

Near a continuous zero-temperature phase transition governed by a quantum critical
point, the critical behavior is expected to obey a standard scaling
formalism\cite{Sondhi}. In particular, for a metal-insulator
transition, the conductivity in the vicinity of the transition
($t \rightarrow t_c, T \rightarrow 0 $)
is expected to scale as:

\begin{equation}
\sigma (t,T) = \sigma_c (T) f[ (t-t_c) T^{ -1/z\nu}].
\label{equation}
\end{equation}
where $ \sigma_c (T) \propto T^{\mu/z\nu} $ is the temperature-dependent
conductivity at $t = t_c$, $\mu$ is the exponent of the zero-temperature
conductivity $ \sigma (t,0) \propto (t-t_c)^\mu $,
$\nu$ is the exponent of the divergent correlation length
$\xi \propto (t-t_c)^{-\nu} $, and $z$ is the dynamical exponent
relating spatial and temporal scales near the critical point
$ \tau  \propto \xi ^z $, with the characteristic temporal scale
at a temperature T given by $ \hbar / k_B T $.

By recasting Equation (1) as

\begin{equation}
\sigma (t,T) = B T^{1/2} [ 1 + A(t-t_c)^\mu/BT^{1/2} ]
\label{equation}
\end{equation}
it is easily seen to be a special case of the scaling form (Eq. (2))
with the identification $ \mu/z\nu = 1/2 $; in conjunction with
the experimentally determined value $\mu = 1$, this implies $ z\nu = 2$.

In contrast, the situation in uncompensated doped semiconductors,
such as Si:P, Si:B and Ge:Ga, appears to be much more complicated,
and there continues to be debate and controversy (see {\em e.g.},
\cite{TPR,Lohneysen,sarachikmott}) concerning the behavior of the conductivity
near the metal-insulator transition.  Far from the transition deep in the 
metallic phase, the conductivity clearly exhibits a $T^{1/2}$ 
dependence\cite{Rosen3} at low temperatures, in agreement with Eq. (1) and 
consistent with perturbative results for a weakly disordered metal
\cite{Altshuler}.  The coefficient $B$ is found to depend weakly on 
dopant concentration deep in the metal, in qualitative agreement with 
theoretical expectations\cite{bhattlee}.
Closer to the transition, however, the dependence of $B$ on
concentration becomes rather marked, and actually changes
sign from negative to positive as the transition is approached.  
The full scaling relation, Eq. (2), is not satisfied if one includes in the 
analysis both negative and positive slopes $B$.  (In fact, if the conductivity 
obeys Eq. (1) near the transition, then scaling 
requires that the coefficient $B$ scale as a power of $(t-t_c)$, so that   
reversals in the sign of $B$ are explicitly excluded).
In the region close to the transition where the temperature coefficient of 
the low-temperature conductivity is positive, even the form of the
temperature dependence of the conductivity is not clearly established: it 
has been reported in different experiments as $\propto T^{1/2}$ and 
$\propto T^{1/3}$.\cite{T1/3}

Very different critical conductivity exponents have been obtained in 
uncompensated Si:P, considered the 
prototypical doped semiconductor.  A value $\mu=0.5$ was found in the 
classic experiments of Paalanen {\it et al.}\cite{paalanen} down to very 
low temperatures (below 5 mK), where uniaxial stress was 
used to tune the transition.  In experiments where the transition 
was approached by reducing the dopant concentration, similar exponents 
near $0.5$ were found in Si:P\cite{sarachikSi:P} as well as a number of 
other uncompensated doped 
semiconductors, including Si:As\cite{castner}, double-doped 
Si:P,As\cite{holcomb} and Ge:Ga\cite{itoh}.  In contrast, 
Stupp {\it et al.}\cite{stupp} 
found $\mu = 1.3$ in Si:P, and Shlimak {\it et al.}\cite{shlimak} deduced $\mu=1$ 
for uncompensated 
transmutation-doped Ge:Sb.  These large 
exponents were based on data in a narrow range of dopant concentration 
near the transition where the coefficient $B$ of Eq. (1) 
is positive.  Using dopant concentration to tune the transition, 
a prior study involving one of the present authors
reported $\mu=0.65$ in Si:B, a material in which the impurity states 
are characterized by an angular momentum $ J= 3/2 $ arising from spin-orbit
coupling characteristic of the valence bands of semiconductors like Si,
and where spin-orbit scattering has been found to be strong\cite{ourSi:B}.

We have recently reported\cite{ourPRL} measurements of the conductivity in 
Si:B in the immediate vicinity of the transition. By applying a
compressive uniaxial stress, $S$, along the [001] direction using a pressure 
cell described elsewhere\cite{bogdanovich}, we have driven a sample of 
Si:B from the metallic phase toward the transition, and mapped out 
the conductivity as a function of applied stress $(S)$ and temperature $(T)$ 
in the range $0.05 K < T < 0.5 K$.  We find that the 
conductivity is described accurately by the scaling form given by Eq. (2) 
(with $t=S$) for a range of stresses which yield conductivities that obey 
Eq. (1) with a constant coefficient $B$. However, the critical conductivity 
exponent is found to be $ \mu \approx 1.6 $, considerably larger 
than the values around $\mu = 0.5 - 0.7 $ reported by many workers,
including that reported earlier for Si:B\cite{ourSi:B}
where the transition was approached by varying the dopant 
concentration.

In this paper, we describe in detail the measurements on the metallic 
side of the transition and compare results obtained on a sample
subjected to uniaxial stress with those obtained earlier for a series of 
unstressed samples in Ref.\onlinecite{ourSi:B}. We are led to the 
surprising conclusion that the two do not agree in detail, suggesting that 
further investigation of the issue of critical behavior in the presence of 
uniaxial stress is warranted. We describe the experimental details and results 
below, followed by a discussion and summary.

\section{Experimental Details and Results}
\label{exdetails}

	A bar-shaped $8.0$x$1.25$x$0.3$ mm$^3$ sample of Si:B was cut with its long 
dimension along the [001] direction.  Relatively small uniaxial stress has a 
pronounced effect on the conductivity of Si:B, driving it initially toward 
more insulating behavior.  A detailed discussion of the effect of stress 
is contained in a companion paper.  The dopant concentration, determined 
from the ratio of the resistivities\cite{bogdanovich} at 300 K and 4.2 K, 
was $4.84$x$10^{18}$ cm$^{-3}$.  Electrical contact was made along four 
thin boron-implanted strips.  Uniaxial compression was applied 
to the sample along the long [001] direction using a pressure cell described 
elsewhere\cite{bogdanovich}.  Four-terminal measurements were taken at 13 Hz (equivalent 
to DC) for different fixed values of uniaxial stress at temperatures between 
0.05 and 0.75 K.  Resistivities were determined from the linear 
region of the I-V curves.

	As discussed earlier, Eq. (1) is expected to be valid 
at low temperatures in the weakly disordered metal
(perturbative regime)\cite{Altshuler,leeramareview}, i. e. not too close to the 
transition. In the 
absence of theoretical predictions very near the 
transition, the conductivity is often fitted to this form everywhere, including 
the critical 
regime. Following this generally accepted procedure,
we plot the conductivity of Si:B
as a function of $T^{1/2}$ for various 
values of the stress $S$ in Fig. 1.  In agreement with experiments where 
dopant concentration is used to tune the transition, the slope $B$ 
of the curves changes from negative to positive with increasing stress 
as the critical value $S_c$ is approached.  However, although the apparent 
straight-line behavior implies the validity of Eq. (1), an equally 
good fit (not shown) is obtained by plotting the data as a function 
of $T^{1/3}$.  This method is therefore not sufficient to distinguish 
between the two functional forms.  

We now present the results of a full scaling analysis of these data 
published elsewhere\cite{ourPRL} and discuss its implications.  
The critical stress for the sample used in our experiments was determined 
to be $S_c = 613$ bar; the temperature dependence at this value of stress, 
{\it i. e.} the critical conductivity, is $\sigma_c (T) \propto T^{0.5}$. 
We rewrite the scaling form, Eq. (2), as:

\begin{equation}
\sigma(S,T) = \sigma(S,0) G[T/(\Delta S/S_c)^{z\nu}]
\label{equation}
\end{equation}
where $\Delta S = (S - S_c) $  and $\sigma(S,0) = \sigma_0 (\Delta S/S_c)^\mu $.
Guided by this version of the scaling form (Eq. (4)), the quantity
$\sigma (S, T)/(\Delta S/S_c)^\mu$ 
is plotted in Fig. 2 (a) as a function of the scaling variable, 
$T/(\Delta S/S_c)^{z\nu}$ with $z\nu=3.2$ and $\mu=1.6$ chosen to yield 
the best data collapse\cite{ourPRL}.  The resulting scaling 
function fully describes the temperature dependence of the conductivity in 
the conducting phase in the vicinity of the transition.  If the 
usual assumption is made that $\mu = \nu$, then the dynamical exponent 
$z = 2$, the same as that found in systems described by Eq. (1), such
as semiconductor-metal mixtures and persistent photoconductors.

To test whether Eq. (1) provides a good description of the conductivity of Si:B 
very near the transition, we replot the same data as a function of  
$[ T/ ( \Delta S/S_c)^{z\nu} ]^{1/2} $ in Fig 2(b).
The data fall nearly on a straight line, indicating that the
temperature dependence of the conductivity of Si:B just on the
metallic side of the metal-insulator transition in the scaling
regime is rather similar to that of metal-semiconductor mixtures and
doped, highly compensated $Al_x Ga_{1-x} As$.  This, in turn,
implies that the $T^{1/2}$ corrections exhibited by the conductivity 
in the perturbative regime of the weakly disordered metal
extend all the way to the critical point\cite{castellani}.
Pronounced failure of scaling occurs if we assume a critical temperature 
dependence in Si:B of $T^{1/3}$ instead of $T^{1/2}$; we are thus able  
to assert that the temperature dependence
of the critical curve and the scaling function are decidedly inconsistent 
with the $T^{1/3}$ dependence that has been found in some other 
materials, such as Ge:Ga\cite{itohpreprint} and Ge:Sb\cite{shlimak}.  
Since Si:B and Ge:Ga are both acceptor 
systems, it would be of
importance to see if similar scaling holds in the latter case, and
whether the critical curve displays similar $T^{1/3}$ dependence.

A best straight line\cite{footnote,note} fit to the data of Fig. 2 (b) yields:
\begin{equation}
\sigma (S, T)/(\Delta S/S_c)^{\mu} = 66 +10.6 [T/(\Delta S/S_c)^{z\nu}]^{1/2}.
\label{eq}
\end{equation}
Rearranging terms and making use of the fact that 
in our case
$\mu = z\nu/2 = 1.6$, this can be written as:
\begin{equation}
\sigma (S, T) = 66 (\Delta S/S_c)^{1.6} + 10.6 T^{1/2}
\label{eq}
\end{equation}
where $\sigma$ is in $(ohm-cm)^{-1}$ and $T$ is in Kelvin.
This is precisely of the form Eq. (1), as stated earlier.

A striking feature of these results is the very large critical 
conductivity exponent $1.6$ compared to the exponent $0.65$ found 
in earlier experiments\cite{ourSi:B} where the transition was approached 
by tuning the dopant concentration. This is further illustrated in Fig. 3, 
which shows the zero-temperature conductivity of the stressed sample 
plotted as a function of $\Delta t/t_c$ on a linear scale compared with 
the conductivity obtained from a series of unstressed samples with varying 
dopant density.  The symbols represent zero-temperature extrapolations 
obtained from the $T^{1/2}$ curves of Fig. 1 and the lower solid 
curve represents the first term on the right of Eq. (6); here the tuning 
parameter $t=S$.  The upper solid curve represents the zero-temperature 
conductivity as a function of dopant concentration taken from Reference 6; 
here the tuning parameter $t=n$.  The difference between the results for 
stressed and unstressed samples is clear and dramatic.

To probe these differences further, we show in Fig. 4 the temperature
dependence of the conductivities of a series of unstressed metallic samples
close to the metal-insulator transition from Reference\onlinecite{ourSi:B}
(shown as open circles) along with the data of Fig. 1.
The two sets of data clearly do not overlap, as might be expected if tuning 
through the transition by varying stress or dopant concentration were equivalent.
Although magnetic field-tuned transitions have long been recognized 
as different and belonging to a different universality class, it has 
generally been assumed that stress-tuned and 
concentration-tuned transitions are equivalent, allowing for 
direct comparisons of the critical behavior and critical exponents.  
The data of Fig. 4 seem to indicate that this is not the case. 
We discuss this point further in the next section.

\section{Discussion and Concluding Remarks}
\label{discussion}

The conductivity data in a metallic sample of Si:B subjected to a
uniaxial stress along the [100] direction shows clear evidence of
scaling with temperature and stress as the metal-insulator transition
is approached, in contrast with most previous data on uncompensated doped
semiconductors. The scaling behavior enables one to determine
with much more confidence the critical behavior at the transition,
$ \sigma_c(T) \propto T^{0.5} $ than is possible from the temperature
dependence of individual samples. However, the scaling yields a much larger 
critical exponent $ \mu \approx 1.6 $ characterizing the zero-temperature 
conductivity, 
$ \sigma(S,0)
\propto (S_c-S)^\mu $,than in the absence of stress.

This large difference naturally raises a number of questions. As stated
earlier, acceptors in semiconductors are characterized by an angular
momentum variable corresponding to $J = 3/2$, and therefore have a four-fold
degeneracy in the unstressed cubic crystal that is lifted in the presence of
uniaxial stress.  However, time reversal symmetry, which is broken in the
presence of magnetic field, is maintained in the presence of stress (the
acceptor state is now two-fold degenerate). Consequently, the change in
univerality class expected in the presence of a magnetic field is not
expected for uniaxial stress.  If, however, the breaking of the four-fold
degeneracy leads to some (as yet unknown) new universality class, this
effect should be easy to confirm experimentally - in Si:P, where there
is no such degeneracy in zero stress, the unstressed and uniaxially
stressed data should not have the large discrepancies seen in Si:B.

A potential source of error in the determination of the critical exponent
is strong nonlinearity in the stress dependence of the critical density.
In both n- and p-doped Si (or Ge), the change in the critical density with
stress can be attributed to a change in the impurity wavefunction\cite{bhatt1}.
In Si:P, the change at low stresses due to mixing of the central-cell split
excited 1s states into the ground state can be calculated\cite{bhatt2}
and shown to be quadratic in $S$, {\em i.e.}, $ n_c(S) = n_c(0) - aS^2 $.
At large stresses, on the other hand, the impurity wavefunction is derived
(for most directions of stress) from the two lowest conduction
band minima, and so the critical density $n_c(S)$ saturates as $ S \rightarrow
\infty $. As a result, $n_c(S)$ is a monotonic function of $S$, with an "S-shaped"
curve, which can be reasonably approximated by a linear curve for small
excursions around a critical value $S_c$, except for very small and very large
$S$, the characteristic $S$ corresponding to the strain given by the central
cell splitting divided by the conduction band deformation potential.

In Si:B, stress initially splits the $ J = 3/2 $ acceptor state linearly, causing 
a much more dramatic dependence on the stress.
A calculation of the acceptor wavefunction at low stress\cite{Chroboczek,Durst}
does not explain this large dependence; instead the predominant effect must come
from the disappearance of the freedom to choose between orbitally distinct
wavefunctions (as in the case of effective mass donors\cite{bhatt3}). For
large compressive stresses along the [001] direction, on the other hand, the
acceptor wavefunctions must be derived predominantly from the light hole
valence band, and therefore $n_c$ is expected to decrease, as the acceptor
wavefunction expands\cite{Chroboczek,Durst}.  Consequently, $n_c(S)$ actually
exhibits a maximum as a function of $S$, so that an appropriately doped
sample of Si:B should exhibit a reentrant metal-insulator-metal transition
as a function of stress.

In the absence of quantitative theory for $n_c(S)$, we base our assumption 
that nonlinearities are not significant on the experimental finding that  
both the stress dependence of the conductivity
of our sample $\sigma(S,T)$ at a high temperature ($T$ = 4.2K), and the
dopant density dependence of the conductivity of a series of closely spaced 
unstressed samples
$\sigma(n,T)$ at $T$=4.2K are {\em linear} in $S$ and $n$ respectively over
the range of the control parameter around the critical value used
in our analysis.  
Further, the critical stress for our sample, $ S_c = 613 $ bar, lies
well away from zero stress (where one might expect some
complications from local strains due to a Jahn-Teller splitting of the acceptor
state) and from the stress corresponding to the maximum resistivity at low $T$
($ S_{max} = 3.5 $ kbar ), and is therefore less likely to be affected by
nonlinearities in $n_c(S)$.  Confirmation of this must await results on
a series of samples with differing values of the critical stress $S_c$.

Another possible source for the unusually large exponent obtained in the current 
experiments is an inhomogeneous distribution of stresses 
resulting in a spread of $\Delta S$'s and a consequent averaging 
over a distribution of conducting paths, some further and some closer to the 
transition.  However, such a distribution might well be expected 
to give rise to measurable deviations from scaling, and the quality 
of the data collapse shown in Fig. 2 is excellent.  

One also needs to consider possible effects associated with anisotropic 
conductivities in uniaxially stressed samples.
For a sample under [001] stress, the conductivity along the stress direction
$\sigma_l(S)$ differs from the conductivities along the transverse [100] and [010]
directions, $\sigma_t(S)$.  Assuming a normal Fermi liquid metallic phase, 
and since the critical stress $S_c$ is nonzero, the conductivity anisotropy

\begin{equation}
\alpha (S) = 3 [ \sigma_l(S) - \sigma_t(S) ] / [\sigma_l(S) + 2 \sigma_t(S)]
\label{equation}
\end{equation}
may be expanded in an analytic Taylor expansion around $S_c$

\begin{equation}
\alpha (S) = \alpha (S_c) + (d \alpha /dS)_{S_c} (S - S_c)
\label{equation}
\end{equation}
which can easily be shown to lead to a subleading correction to the
conductivity onset when measured in any direction, {\em i.e.},
if we take:

\begin{equation}
\sigma_{tr} (S) = [ \sigma_l (S) + 2 \sigma_t (S) ]/3 = \sigma_0 [(S_c-S)/S_c]^\mu
\label{equation}
\end{equation}
we obtain

\begin{equation}
\sigma_l(S) , \sigma_t(S) \propto (S_c-S)^\mu [ 1 + O(S_c-S) ]
\label{equation}
\end{equation}
where the coefficient of the term of order $(S_c - S)$ in the square
brackets will be proportional to $ ( d \alpha /dS)_{S_c} $.

The anisotropy also affects the comparison between unstressed and
uniaxially stressed samples shown in Fig. 4.  In particular, one expects 
to be able to compare the stress dependence of the angle-averaged
value $\sigma_{tr} (S) $ to the concentration dependence of $\sigma (n)$
of the unstressed (cubic) samples.  Consequently, the longitudinal 
conductivities $\sigma_l (S) $ for the uniaxially
stressed samples (closed circles in Fig. 4) should be divided by a
stress-dependent anisotropy factor $[1 + 2 \alpha (S) /3 ]$ when comparing with 
the unstressed samples.  (In providing a direct comparison in Fig. 4, we have 
assumed that $\alpha $ is small at the stresses applied.)
To test this, we multiplied each of the
unstressed curves (open circles) by an arbitrary factor chosen to make 
it coincide with corresponding curves for the stressed samples.  Our best 
attempt, shown in Fig. 5, requires rather large anisotropy values $\alpha$;  
moreover, the $\alpha$'s (listed in the caption to Fig. 5) are unphysical: they are 
nonmonotonic functions of the stress, and decrease with increasing stress in 
the critical region.  We therefore conclude that the difference between the 
temperature dependence of the uniaxially stressed and unstressed samples is 
intrinsic and not due to effects associated with anisotropic conductivities.  
Conclusive proof would require measurements of the conductivities in both 
the longitudinal and transverse directions in the presence of stress.

It should be noted that the scaling is found to hold only in a relatively
small window of metallic conductivities for control parameter values
rather close to the critical value. The much smaller exponent, $ \mu \approx
0.5 - 0.7 $, is derived from data over a much wider range. There has been much
debate about the unusually small correlation length exponent $\nu$ that such
a small $\mu$ implies, and possible violation of the bound
derived for disordered systems $ \nu \geq (2/d) $\cite{chayes}.  It is not 
clear whether such systems become inhomogeneous at long length scales and 
are then governed by percolation near the transition\cite{zimanyi}.  Such a scenario would  
offer the attractive possibility of reconciling the many different results 
found in Si:P and Si:B.  We point out that reports
of large conductivity exponents\cite{stupp,ourPRL} are confined to a
region very close to the critical value of the tuning parameter, where percolation 
may well result from such inhomogeneities.  
Further, the observed conductivity exponent is close to that expected
for classical percolation in three dimensions\cite{percolation}.  Finally,
this might account for earlier observations\cite{paalanen} of
differing conductivities in different samples very close to the metal insulator
transition; the percolative paths could be rather sensitive to
precise details of dopant distribution, and lead to nonuniversal amplitudes
especially in a crossover region.

\section{summary}
\label{summary}

In summary, we have used compressive uniaxial stress applied along the 
[001] direction to approach the metal-insulator transition from the metallic 
side in Si:B.  The conductivity scales with stress and temperature over the 
narrow range within which Eq. (1) is obeyed with a constant coefficient 
$B$. The temperature dependence of the conductivity at the critical
value of the tuning parameter (uniaxial stress in our case) is found
to be proportional to $T^{0.5}$. The critical exponent characterizing
the onset of the zero-temperature conductivity is found to be $\mu = 1.6$,
considerably larger than the exponent found in experiments where the transition was 
approached by reducing dopant concentration.  The temperature dependence 
of the conductivity is qualitatively and quantitatively different for 
stressed and unstressed Si:B, however, suggesting that a direct 
comparison of the critical exponents is not possible.  Our data call for a 
systematic study of the stress tuned transition in other donor and
acceptor systems, as well as for critical reexamination of the 
assumption that stress-driven and concentration-driven metal-insulator 
transitions are equivalent for all doped semiconductors.

\section{Acknowledgments}

We are grateful to Jonathan Friedman, D. Simonian and S. V. Kravchenko for their
participation in some phases of these experiments.  We acknowledge 
valuable experimental contributions by L. Walkowicz.  
Our heartfelt 
thanks go to G. A. Thomas for his generous support and expert advice, help and 
interest throughout this project.  We thank T. F. Rosenbaum, M. 
Paalanen, E. Smith and S. Han for valuable experimental advice and the loan of equipment, and F. 
Pollak for useful suggestions 
and some samples.  M. P. S. thanks G. Kotliar and D. 
Belitz for numerous discussions, and John Davies for several discussions 
regarding the possible effects of inhomogeneous stress distributions.  
This work was supported by the 
US Department of Energy Grant No.~DE-FG02-84ER45153.  R. N. B. was 
supported by NSF grant No. DMR-9400362 and DMR-9809483.

\begin{figure}
\caption{Conductivity of Si:B versus $T^{1/2}$ for different values of uniaxial 
stress that place the sample in the metallic phase.  The critical stress for 
this sample is 613 bar.}
\label{fig1}
\end{figure}
\begin{figure}
\caption{ (a) $\sigma(S,T)/(\Delta S/S_c)^\mu$ as a function of the scaling 
variable $T/(\Delta S/S_c)^{z\nu}$ with $\mu = 1.6$ and $z\nu = 3.2$ 
determined in ref. 13. Here $\Delta S = (S - S_c)$, where $S_c$ is the critical 
stress.  (b) $\sigma(S,T)/(\Delta S/S_c)^\mu$ versus 
$[T/(\Delta S/S_c)^{z\nu}]^{1/2}$.}
\label{fig2}
\end{figure}
\begin{figure}
\caption{The zero-temperature conductivity, $\sigma(T \rightarrow 0)$, versus 
$|\Delta t|/t_c$.  The lower curve represents Eq. (3) with the parameters 
$\sigma_0 = 66$ (ohm-cm)$^{-1}$ and $\mu=1.6$ found for the stress-driven transition (for 
which $t=S$); the symbols 
represent $T=0$ extrapolations obtained from the 
temperature-dependent curves of Fig. 1.  The upper curve, taken from 
ref. 13, shows the critical behavior found earlier in experiments where 
the transition was approached by reducing the dopant concentration, $n=t$.  Here 
$\sigma_0 = 152$ (ohm-cm)$^{-1}$ and $\mu=0.65$}
\label{fig3}
\end{figure}
\begin{figure}
\caption{The conductivity of stressed and unstressed Si:B plotted as a 
function of $T^{1/2}$. Closed circles denote data for a sample under stress 
and open circles indicate data for unstressed Si:B with different dopant 
concentrations as labelled.} 
\label{fig4}
\end{figure}
\begin{figure}
\caption{Closed symbols denote the conductivity of stressed Si:B 
plotted as a function of $T^{1/2}$. The open square symbols 
represent the conductivity of four  
unstressed samples multiplied by an arbitrary factor $K= [1 + 2 \alpha (S) /3 ]$ 
chosen to make them coincide with corresponding curves for the stressed samples, 
as follows: curves $1, 2, 3,$ and $4$ denote $n=4.11, 4.20, 4.30$, and $4.38 \times 
10^{18}$ cm$^{-3}$ with multiplicative factors $K_1=1.18, K_2=1.27, 
K_3=1.40$, and $K_4=1.32$, respectively; clearly, the unstressed curve at $n=4.84  \times 
10^{18}$ cm$^{-3}$ corresponds to $K=1$} 
\label{fig5}
\end{figure}

\begin{references}
\bibitem{footnote}Data for the stress closest to the critical value $S_c$ were 
not included in the fit.  The slope B for the $S= 583$-bar curve is smaller and 
atypical, due possibly to inhomogeneous distribution of stress resulting in 
insulating behavior for some regions of the sample.
\bibitem{note} A somewhat better fit is obtained using $T^{0.47}$.  
This represents the uncertainty associated with the present analysis.
\end{references}
\end {document}
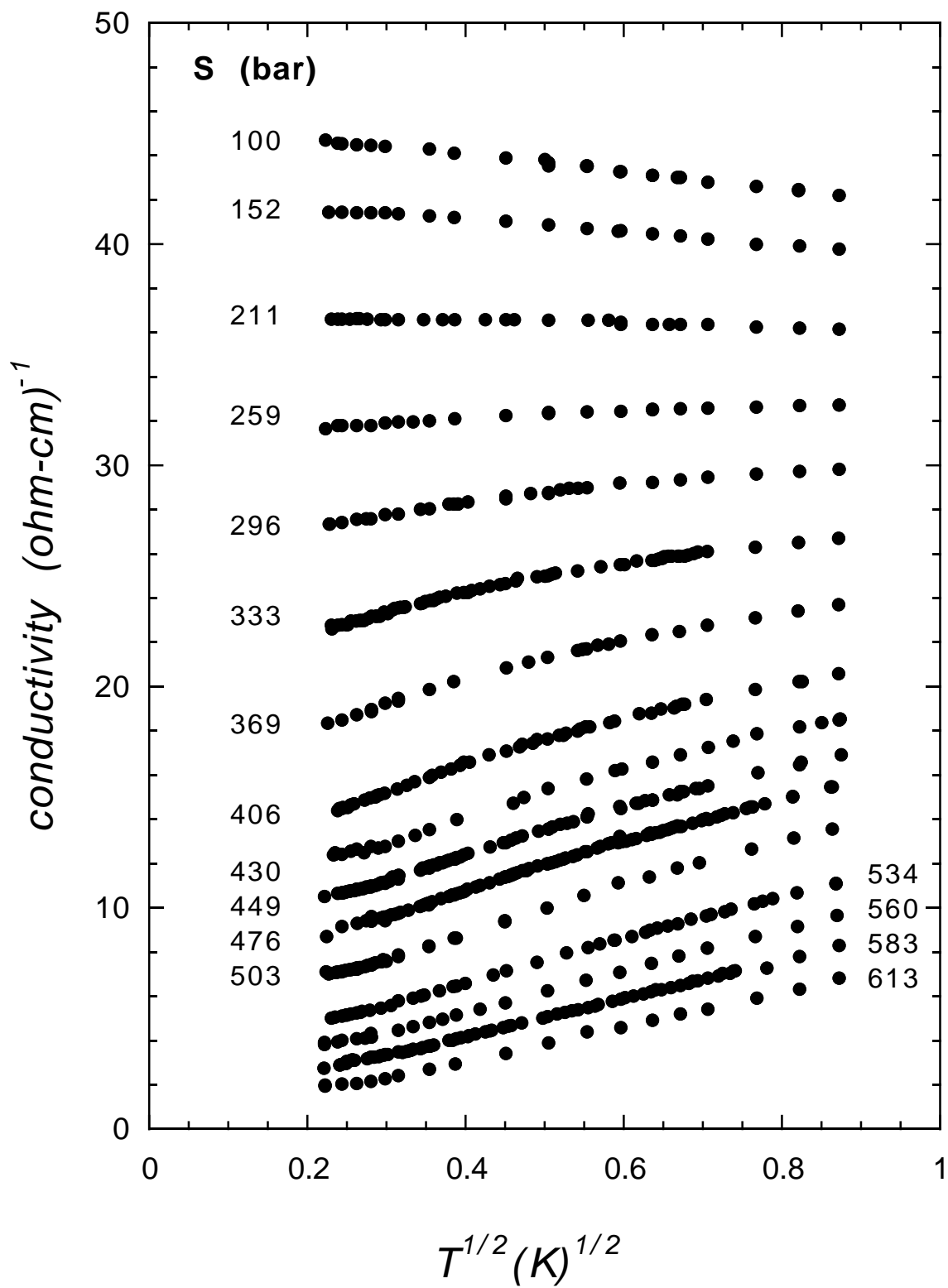
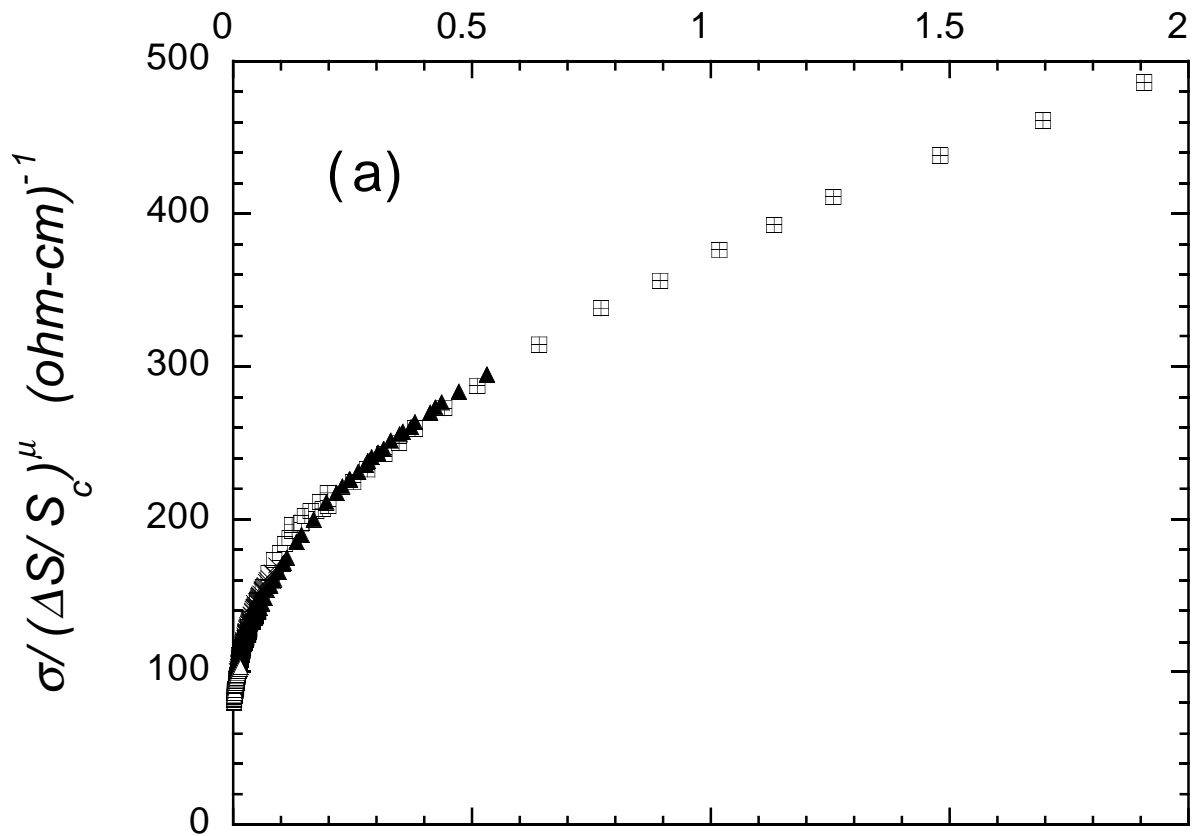
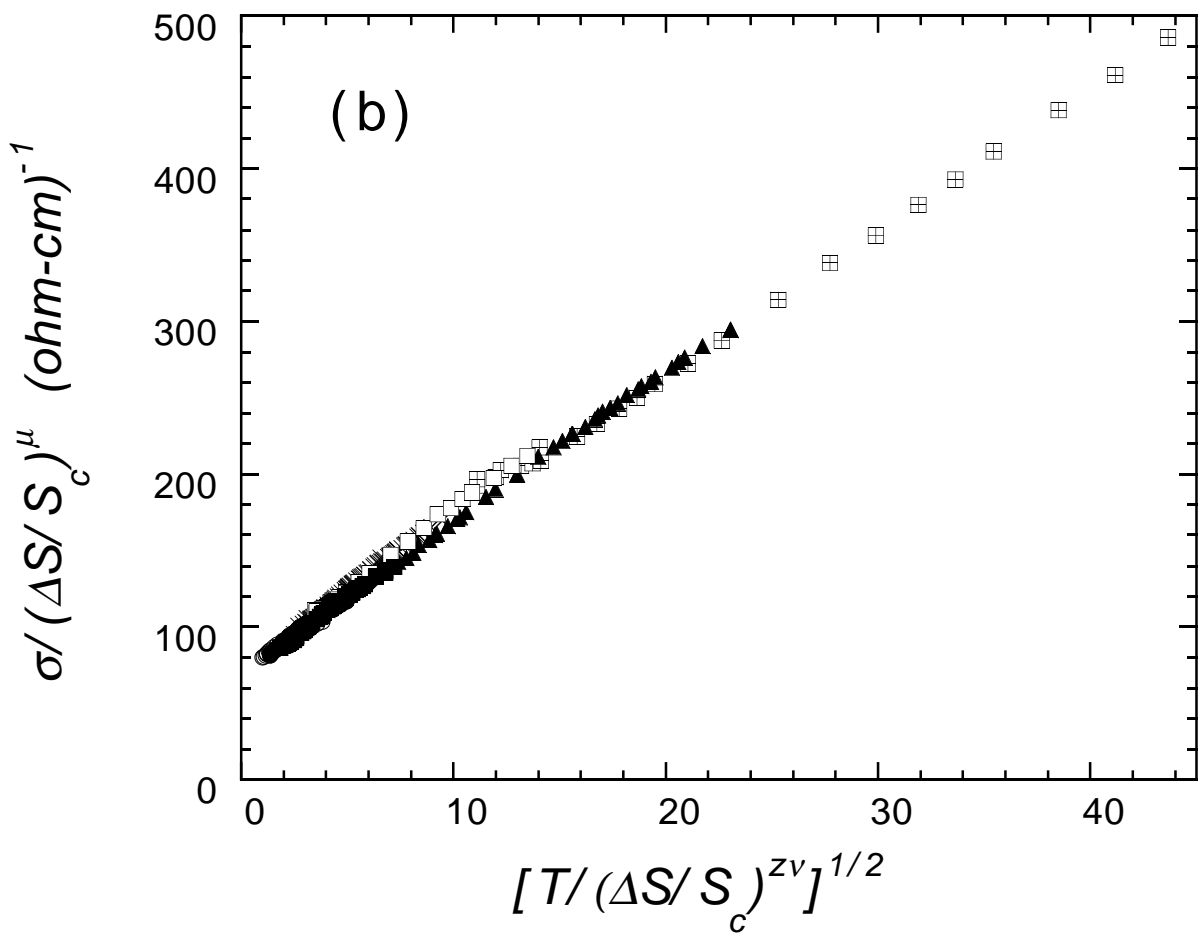
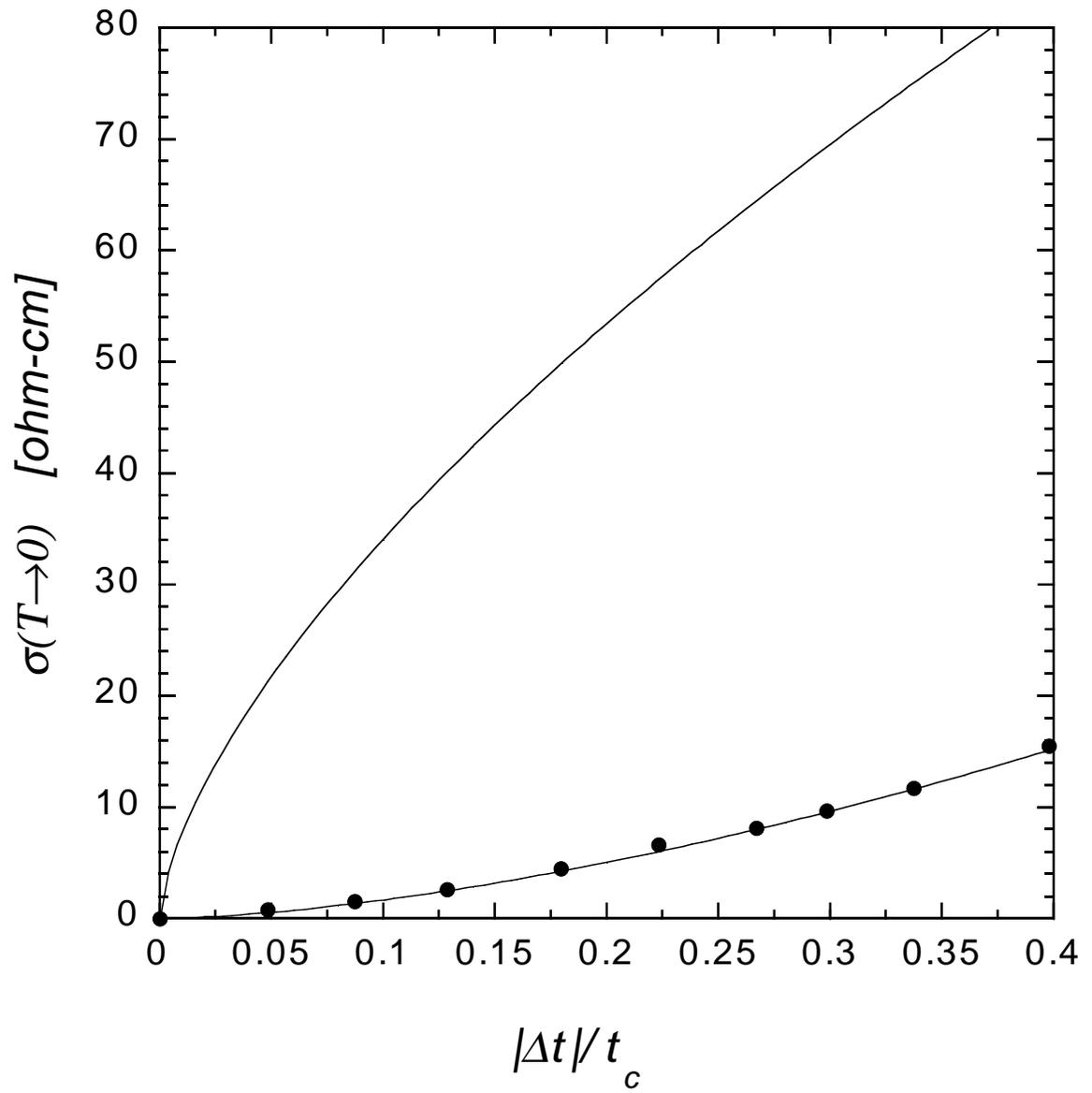
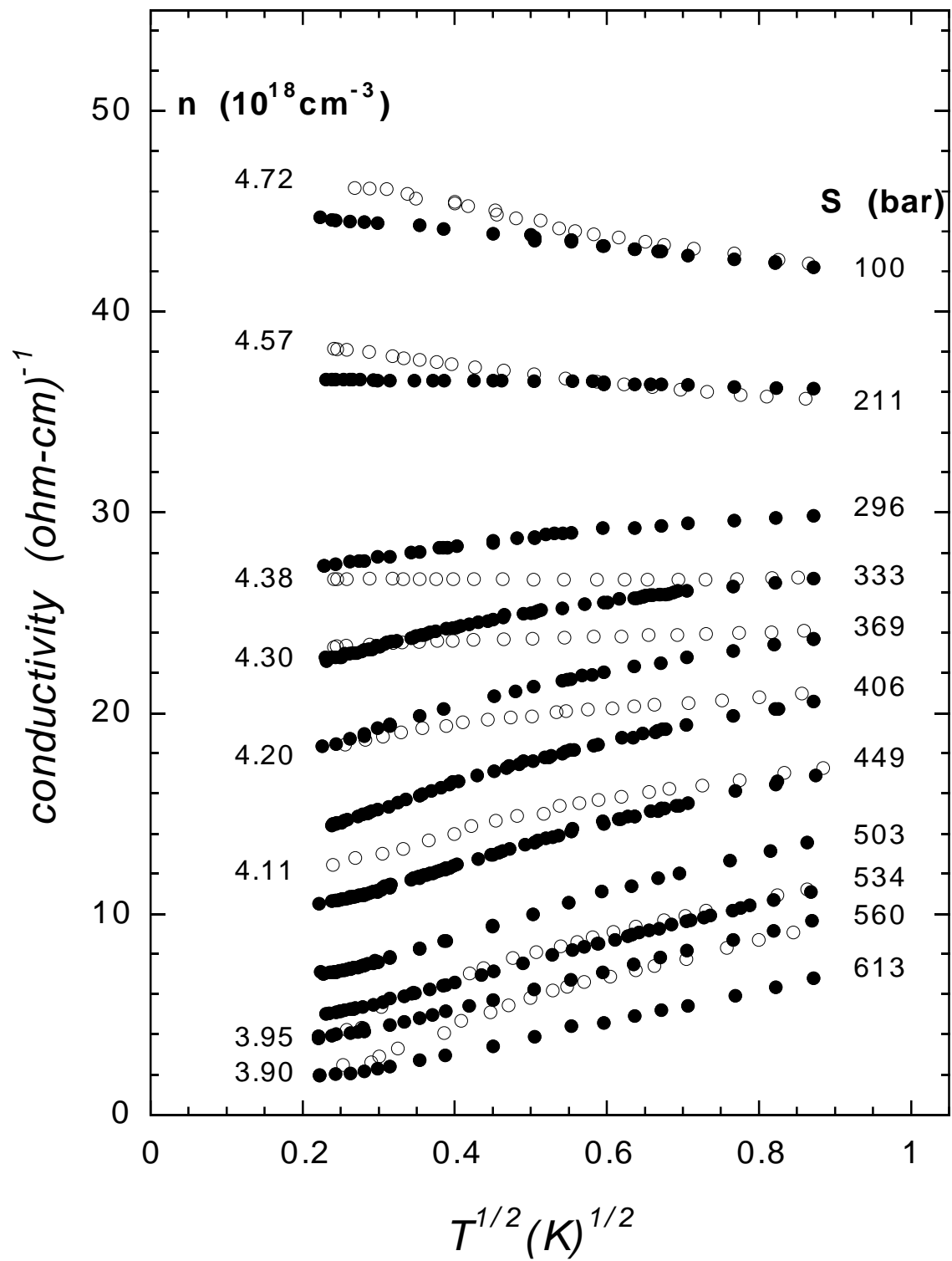
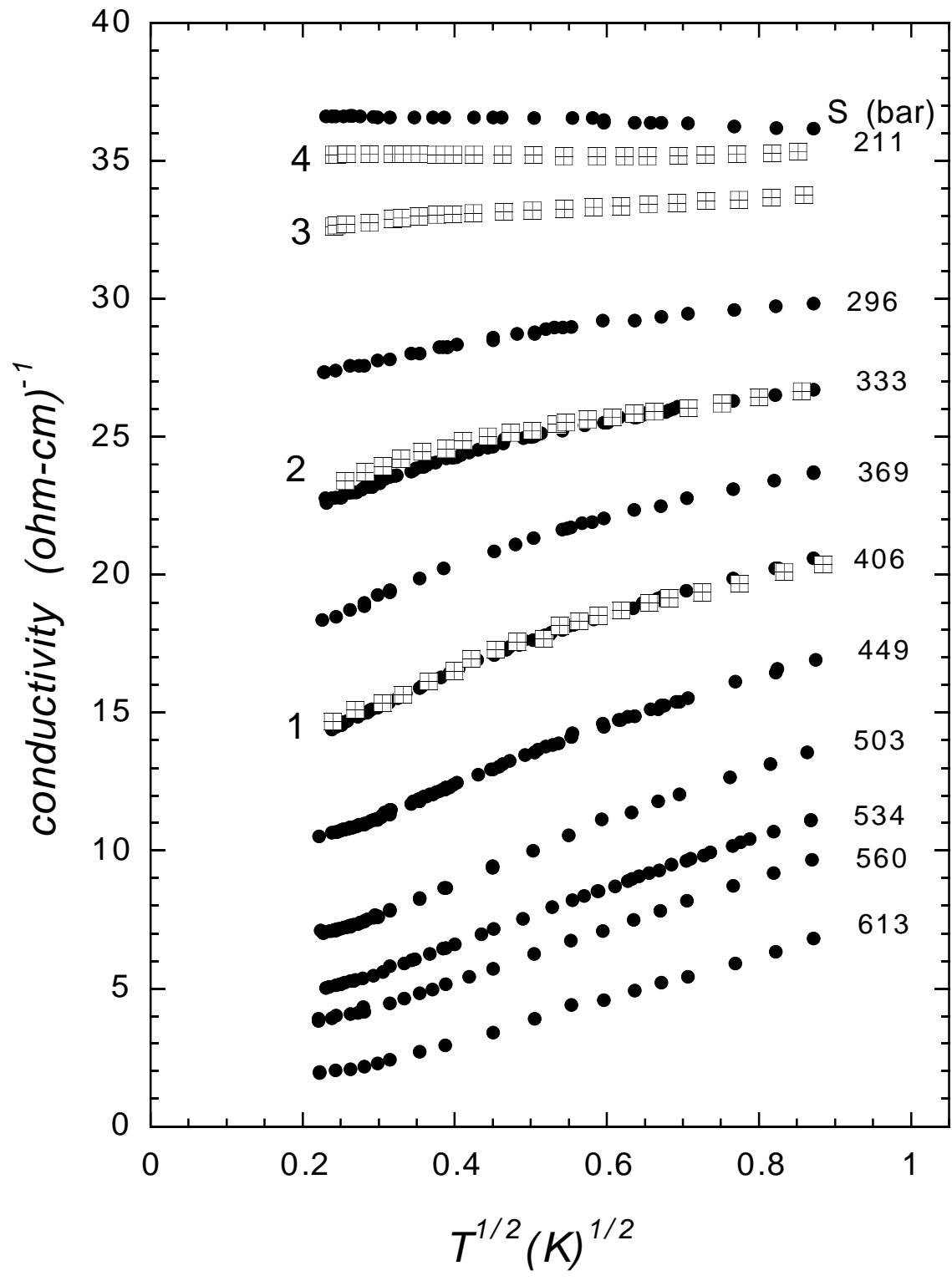
